\documentclass[doublecol]{epl2} 
\usepackage{amssymb}

\title{Global generalized synchronization in networks of different time-delay systems}

\author{D. V. Senthilkumar\inst{1}, R.~Suresh\inst{2}, M. Lakshmanan\inst{2} \and J. Kurths\inst{1,3,4}}

\institute{                    
  \inst{1} Potsdam Institute for Climate Impact Research, 14473 Potsdam, Germany\\
  \inst{2} Centre for Nonlinear Dynamics, School of Physics, Bharathidasan University, Tiruchirapalli 620 024, India\\
  \inst{3} Institute of Physics, Humboldt University, 12489 Berlin, Germany\\
  \inst{4} Institute for Complex Systems and Mathematical Biology, University of Aberdeen, Aberdeen AB24 3UE, United Kingdom}

\pacs{05.45.Xt}{Synchronization; coupled oscillators}
\pacs{05.45.Pq}{Numerical simulations of chaotic systems}
\pacs{05.45.-a}{Nonlinear dynamics and chaos}

\abstract{
We show that global generalized synchronization (GS) exists in structurally different time-delay
systems, even with different orders, with  quite different fractal 
(Kaplan-Yorke) dimensions, which emerges via
partial GS in symmetrically coupled  regular networks.
We find that  there exists a smooth transformation  in such systems,
which maps them to a common GS manifold  as corroborated by their maximal transverse Lyapunov exponent.
In addition, an analytical stability condition using the Krasvoskii-Lyapunov theory
is deduced. This phenomenon of GS in strongly distinct systems opens a new way for an
effective control of pathological synchronous activity by means of extremely small
perturbations to appropriate variables in the synchronization manifold.}

\begin{document}

\maketitle

Synchronization is a ubiquitous nonlinear phenomenon serving as a 
platform for information processing in diverse
natural and man made systems~\cite{asp2001,ml2011}. It 
has been investigated mainly in identical systems and in
systems with parameter mismatch, with rare exceptions of distinctly (structurally)
nonidentical systems~\cite{asp2001,ml2011}. However, in reality very often synchronization emerges 
in distinctly nonidentical systems such as
respiratory arrhythmia between cardiovascular and respiratory systems~\cite{csmgr1998},
visual and motor systems~\cite{sff1998},
paced maternal breathing on fetal~\cite{pvldg2009},
different populations of species~\cite{bbah1999,reagr2006},
in epidemics~\cite{btgonb2000}, in climatology~\cite{ksat2011}, and many more.
Considering the coherent coordination of living systems involving multiple organs such
as brain, heart, lungs, limbs, etc., or machines consisting of 
distinct parts, cooperative evolution of distinct and often
time-delayed systems is essential and challenging.

Among various kinds of synchronization admitted by coupled nonlinear dynamical
systems~\cite{asp2001,ml2011}, the intricate phenomenon of generalized synchronization 
(GS) refers to (static) functional relationship between interacting systems~\cite{lkup1996,nfrmms1995,rb1998,kp1996,zzxw2002,ychyth2008,sg2009,jc2009,ah2010,sg2010,ys2009,oimaak2012,emskas2009,sa2013,sbdlv2000,zzgh2000}.
While the phenomenon of GS has been well understood in unidirectionally 
coupled systems~\cite{lkup1996,nfrmms1995,rb1998,kp1996}, it remain still in its infancy 
in bidirectionally  coupled systems
and in particular there
exists only very limited results on GS, even in systems with parameter
mismatches~\cite{zzxw2002,ychyth2008,sg2009,jc2009,ah2010,sg2010,ys2009,oimaak2012,emskas2009,sa2013}, and particularly in structurally
different (nonidentical) systems with different fractal dimensions~\cite{sbdlv2000}. 
Thus, in general, the notion of GS in mutually coupled
systems needs much indepth investigation and in particular
in distinctly different systems with different fractal dimensions involving time-delay. 
Indeed, recent investigations have revealed that
GS is essentially more likely to occur in complex networks (even with 
identical nodes)~\cite{ychyth2008} due to the large
heterogeneity (degree distribution) of many natural networks~\cite{oimaak2012}.

It is important to recall that the above mentioned studies~\cite{csmgr1998,sff1998,pvldg2009,bbah1999,reagr2006,btgonb2000,ksat2011}
have demonstrated only  phase synchronization (PS)  among such distinctly different complex 
systems, while the natural choice of GS in them has been largely neglected,
except for the important study of Zheng etal~\cite{zzxw2002} on low-dimensional
systems without delay and without a substantial difference in their fractal dimension.
Further, depending on the relation between PS and GS~\cite{fn1} in such systems,
which remains unclear, our understanding on their  evolutionary
mechanism, dynamical and functional behavior may need to be reinvestigated.
Controlling pathological synchronous activity and inducing coherent
coordination in paralysed systems may be effectively done upon understanding
the emergence of a common GS manifold in such systems.
Furthermore, in a more general scenerio of
networks of distinctly nonidentical time-delay systems with different fractal dimensions, it remains
unclear whether there exists a transformation (map) that maps the individual systems to a
common GS manifold despite disparity in their degree of complexity. 

In line with the above discussions, in this Letter we will provide
a substantial extension of the formulation of GS to
mutually coupled distinctly different delay systems~\cite{fn2}, which also holds for
systems without delay. 
Based on our generalized formulation, we will specifically demonstrate 
the existence of {\it global GS}  via {\it partial GS} in  
symmetrically coupled networks,  which even consist of
distinctly different time-delay systems (Mackey-Glass~\cite{mcmlg1977}, 
piecewise linear~\cite{dvs2006}, threshold nonlinearity~\cite{dvs2010} and
Ikeda~\cite{kihd1980}). 
It is important to note that such a phenomenon
also occurs among delay systems of different orders, namely, Ikeda and
a second order Hopfield neural network~\cite{jjh1982} as well as Mackey-Glass and
a third order plankton model~\cite{era1998} with multiple delays. 
It is surprising that there exists a common GS manifold even in an ensemble of
distinctly different time-delay systems  to achieve {\it global GS} 
in four different network (array, ring, star and all-to-all) configurations, 
i.e, there exists a function (smooth map)
for each system, even with different fractal dimensions, in a network which 
maps them to the common GS manifold.
We calculate the maximal transverse Lyapunov exponent (MTLE) to evaluate the asymptotic
stability of the complete synchronization (CS) manifold of each of the main systems 
with their corresponding auxiliary systems,
which in turn asserts the stability of the GS manifold between the main systems.
Further, we will
also estimate cross correlation (CC) and correlation of probability of recurrence (CPR)~\cite{nmmcr2007}
to establish relations between GS and PS. 

Kocarev and Parlitz~\cite{lkup1996} formulated that  GS in drive $\vect{x}$
and response $\vect{y}$ configuration occurs only if the response  is 
asymptotically stable, i.e.  $\forall~\vect{y_i}(0), \vect{x}(0)$ in the basin of
synchronization manifold  $\lim_{t \to \infty}\vert\vert \vect{y}(t,\vect{x}(0),\vect{y}_1(0))
-\vect{y}(t,\vect{x}(0),\vect{y}_2(0))\vert\vert=0$, a mathematical formulation of
the concept of auxiliary system approach~\cite{nfrmms1995}. 
Now we give a substantial extension of this formulation to
{\it mutually} coupled different dynamical systems with delay represented as
\begin{equation}
\vect{\dot{x}}=\vect{f}(\vect{x},\vect{x}_\tau,\vect{u}),\qquad
\vect{\dot{y}}=\vect{g}(\vect{y},\vect{y}_\tau,\vect{v}),\qquad \vect{f}\ne \vect{g}
\label{mc}
\end{equation}
where $\vect{x}, \vect{x}_\tau \in\vect{R}^n,  
\vect{y}, \vect{y}_\tau\in\vect{R}^m$, $\tau\in\vect{R}$ and  $\vect{u},\vect{v}\in\vect{R}^k, k\le m,n$.
$u_i=-v_i=h_i\left(\vect{x}(t,\vect{x}_0),\vect{y}(t,\vect{y}_0)\right)$ 
correspond to 
the driving signals. System (\ref{mc}) is in GS if there exists a transformation
$\vect{H}:(\vect{R}^n, \vect{R}^m)\to~\subset\vect{R}^n\times \vect{R}^m$. That is, there may exist a set of transformations $H$ that maps a given $\vect{x},\vect{x}_{\tau}\in \vect{R}^n$
and $\vect{y},\vect{y}_{\tau}\in \vect{R}^m$ to different subspaces of $\vect{R}^n \times \vect{R}^m$.
This implies that the synchronization manifold
$M=\{(\vect{x},\vect{y}): \vect{H}(\vect{x},\vect{y})=0\}$  is such that 
$\forall~\vect{x}(\hat\tau), \vect{y}(\hat\tau), \hat\tau\in\left[-\tau,0\right]$,
which lies within the subset of the basin of attraction $B=B_{\vect{x}}\times B_{\vect{y}}\subset\vect{R}^{n}\times\vect{R}^{m}$,
approaches $M\subset B$, so that $M$ is an attracting manifold. Here $B_{\vect{x}}$ and $B_{\vect{y}}$
are the basins of attractions of systems $x$ and $y$, respectively.

%
\begin{figure}
\centering
\includegraphics[width=0.6\columnwidth]{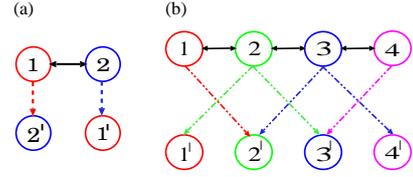}
\caption{\label{fig1}(color on-line) Schematic diagram of the auxiliary system approach for 
mutually coupled time-delay systems for the case of a linear array for (a) $N=2$ and (b) $N=4$.}
\end{figure}
Thus, GS exists in system (\ref{mc}) only if both coupled systems are 
asymptotically stable such that $\forall~(\vect{x}_i(\hat\tau),
\vect{y}_i(\hat\tau)), \hat\tau\in\left[-\tau,0\right]\subset B,~i=1,2$,
$\lim_{t \to \infty}\vert\vert \vect{y}(t,\vect{x}_1(\hat\tau),\vect{y}_1(\hat\tau))
-\vect{y}(t,\vect{x}_1(\hat\tau),\vect{y}_2(\hat\tau))\vert\vert=0$ and 
$\lim_{t \to \infty}\vert\vert \vect{x}(t,\vect{x}_1(\hat\tau),\vect{y}_1(\hat\tau))
-\vect{x}(t,\vect{x}_2(\hat\tau),\vect{y}_1(\hat\tau))\vert\vert=0$.
This is a mathematical formulation of the auxiliary system approach to system (\ref{mc}),
whose schematic diagrams with the main and auxiliary
systems for the case of a linear array  are depicted in Figs.~\ref{fig1}a
and ~\ref{fig1}b, respectively,  which are similar in approach to 
that of~\cite{zzxw2002} concerned with low-dimensional systems without delay. 
It is to be noted that the asymptotic stability of synchronization of
the auxiliary systems and their corresponding original systems holds good only when there
is no subharmonic entrainment of unstable periodic orbits~\cite{uplj1997}.
Therefore trajectories of (\ref{mc}) starting from $B$ 
asymptotically reach $M$ defined by the transformation function $\vect{H}(\vect{x},\vect{y})$,
which can be smooth if the systems (\ref{mc}) uniformly converge (otherwise nonsmooth),
i.e. their local Lyapunov exponents (LLEs) are always negative, to $M$~\cite{lmptlc1997}.


Now, we will demonstrate the existence of GS in symmetrically coupled arbitrary network 
of distinctly nonidentical time-delay systems with different fractal 
dimensions using the above formalism. The dynamics of the $i$th node in the network is
represented as 
\begin{equation}
\dot{\vect{x}}_i= -\alpha_i \vect{x}_i(t)+\beta_i \vect{f}_{i}(\vect{x}_i(t-\tau_{i}))
-\varepsilon \sum_{j=1}^N G_{ij} \vect{x}_{j}, 
\end{equation}
where $i=1,...,N,~N$ is the number of nodes in the network, 
$\alpha_i$ and $\beta_i$ are constant parameters, 
$\tau_i$ are the delay times, $\vect{f}_{i}$ is the nonlinear vector function of 
$i$th node 
and $G$ is an  Laplacian matrix, 
determining the topology of the arbitrary network.
To determine the asymptotic stability of each of nodes in
the network, we define an idential (auxiliary) network (starting from different 
initial conditions) with node $i$ represented
as (see Fig.~\ref{fig1}) 
%
\begin{equation}
\dot{\vect{x}^\prime}_i= -\alpha_i \vect{x}^\prime_i(t)+\beta_i \vect{f}^\prime_{i}(\vect{x}_i^\prime(t-\tau_{i}))-\varepsilon \sum_{j=1}^N G_{ij}(\delta_{ij}(-\vect{x}_{j})+\vect{x}_{j}).
\label{eqn1}
\end{equation}
\begin{figure}
\centering
\includegraphics[width=1.0\columnwidth]{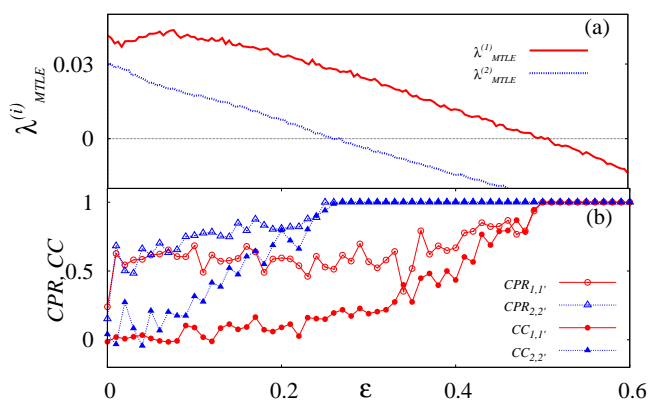}
\caption{\label{mgth_2c}(color on-line) (a) MTLEs and 
(b) CC, CPR of the main and auxiliary systems  of
coupled MG-PWL systems.}
\end{figure}
%
%
%

First, we consider $N=2$ mutually coupled distinctly nonidentical time-delay systems
with the nonlinear  function 
\begin{equation}
f_{1}(x) = \frac{x_{1}(t-\tau_{1})}{(1+(x_{1}(t-\tau_{1})^{10}))},
\end{equation}
of the Mackey-Glass (MG)~\cite{mcmlg1977} system and the piecewise linear (PWL) function
\begin{eqnarray}
f_{2}(x)=
\left\{
\begin{array}{cc}
0, &  x \leq -4/3  \\
            -1.5x-2,&  -4/3 < x \leq -0.8 \\
           x,&  -0.8 < x \leq 0.8 \\
            -1.5x+2,&  -0.8 < x \leq 4/3 \\
0, &  x > 4/3
         \end{array} \right.
\label{eqn3a}
\end{eqnarray}
of the PWL time-delay system~\cite{dvs2006}, respectively.
%
%
%
The parameters of the MG systems are chosen as $\alpha_{1}=0.5$,
$\beta_{1}=1.0$, and $\tau_{1}=8.5$ and for the PWL systems 
we choose $\alpha_{2}=1.0$,  $\beta_{2}=1.2$ and $\tau_{2}=10.0$, 
exhibiting hyperchaotic attractors with two ($D_{KY}=2.957$) and three ($D_{KY}=4.414$) positive Lyapunov 
exponents (LEs)~\cite{ml2011},  respectively.

At a first sight, one may think that mutually interacting systems 
reach a common synchronization manifold simultaneously for a certain 
critical value of $\varepsilon$. But because of the distinctly different coupled systems
with different fractal dimensions, 
one of the systems first reaches a  GS manifold for a lower value of $\varepsilon^{(1)}_{c}$, 
while the other system remains in a desynchronized state, which we refer to as a {\it partial GS}.
With further increase of $\varepsilon$ the other system
also converges at a different critical $\varepsilon^{(2)}_{c}$ 
to the common GS manifold achieving a {\it global GS}.
Further, one may also expect that systems with lower dynamical complexity will
converge to GS manifold first and then the higher one in the order of their degree of
complexity (here in terms of the number of positive LEs). On the contrary, the PWL system
with three positive LEs reaches the GS manifold first at $\varepsilon^{(2)}_c\approx 0.26$ as
indicated by the changes in the sign of the $\lambda^{(2)}_{MTLE}$  
between  $2$ and $2^\prime$ in Fig.~\ref{mgth_2c}a and then
the MG system with only two positive LEs smoothly converges to the GS manifold at 
further larger $\varepsilon^{(1)}_c\approx 0.5$, where $\lambda^{(1)}_{MTLE}<0$
as shown in Fig.~\ref{mgth_2c}a confirming the emergence of a {\it global GS} in distinctly
nonidentical time-delay systems.

The emergence and the transition from partial to global GS is also characterized
by the correlation coefficient (CC)~\cite{fn3}.
If the two systems are in CS state the $CC=1$, 
otherwise $CC<1$. 
Further, the existence of PS in highly non-phase-coherent hyperchaotic attractors 
of time-delay systems are characterized by the  value of the correlation of probability of 
recurrence (CPR $\approx 1$)~\cite{nmmcr2007, dvs2006}.

In the absence of coupling ($\varepsilon=0.0$), $CC_{1,1^{\prime}}$ and $CC_{2,2^{\prime}}$ and
$CPR_{1,1^{\prime}}$ and $CPR_{2,2^{\prime}}$ (\ref{mgth_2c}b) are nearly
zero and both $\lambda^{(1)}_{MTLE}$ and $\lambda^{(2)}_{MTLE}>0$,
indicating the absence of CS (GS) between the main and auxiliary systems. 
If we increase the coupling strength, $CC_{2,2^{\prime}}$ and  $CPR_{2,2^{\prime}}$
start to increase towards unity and
at $\varepsilon^{(1)}_{c}\approx 0.26$, $CC_{2,2^{\prime}}=1$ ($CPR_{2,2^{\prime}}=1$), where $\lambda^{(2)}_{MTLE}<0$ ,
which confirms the onset of GS (PS) in the PWL system while the MG system continues to
remain in a desynchronized state ($CC_{2,2^{\prime}}\approx 0.2$ and $\lambda^{(1)}_{MTLE}>0$).
 Further, if we increase the coupling 
strength to $\varepsilon^{(2)}_{c}\approx 0.5$, a global GS occurs where 
both $CC_{1,1^{\prime}}$ and $CC_{2,2^{\prime}}$ become unity 
and while both $\lambda^{(1)}_{MTLE}$ and $\lambda^{(2)}_{MTLE}<0$.
It is also to be noted that both $CPR_{1,1^{\prime}}$ and $CPR_{2,2^{\prime}}$ are 
also in agreement with their $CC$ confirming the existence of GS and PS together.

\begin{figure*}
\centering
\includegraphics[width=1.8\columnwidth]{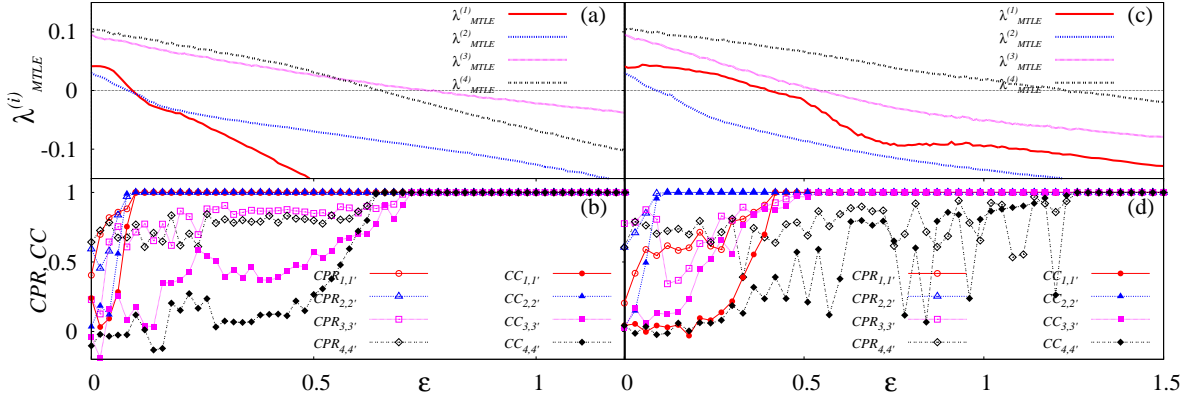}
\caption{\label{4cring}(color on-line) 
 (a) and (c) MTLEs and 
(b) and (d) CC, CPR of the main and auxiliary systems
of four distinctly nonidentical time-delay systems with ring and array
configurations, respectively, as a function of $\varepsilon$.}
\end{figure*}
%
%
We have also analytically investigated the existence of partial and global GS using the 
Krasovskii-Lyapunov theory. For this purpose, we consider the difference in the state
variables of the main and auxiliary systems ($\Delta_{i}=x_{i}-x^{\prime}_{i}$). 
For small values of $\Delta_{i}$,
the evolution equation for the CS manifold for the Mackey-Glass time-delay systems 
($\Delta_{1}=x_{1}-x^{\prime}_{1}$) can be written as 
%
\begin{equation}
\dot{\Delta_{1}}=-(\alpha_{1}+\varepsilon)\Delta_{1}+\beta_{1} [f^{\prime}_{1}(x_{\tau_{1}})] \Delta_{1\tau_{1}} 
\label{eqn1a}
\end{equation}
where $x_{\tau_{1}}=x(t-\tau_{1})$ and $\Delta_{1\tau_{1}}=\Delta_{1}(t-\tau_{1})$. 
The CS manifold is locally attracting if the origin of $\dot{\Delta}_{1}$ is stable. 
Then, a continuous positive definite Lyapunov functional can be defined 
in the form 
%
\begin{equation}
V(t)=\frac{1}{2}\Delta^{2}_{1}+\mu \int_{-\tau}^{0} \Delta^{2}_{1}(t+\theta)d\theta, \qquad V(0)=0,
\label{eqn1b}
\end{equation} 
where $\mu$ is an arbitrary positive parameter ($\mu>0$). The Lyapunov function $V(t)$
approaches zero as $\Delta_{1} \rightarrow 0$. The derivative of $V(t)$
along the trajectory of the CS manifold  should be negative for the 
stability of the CS manifold $\Delta_{1}=0$. 
This requirement results in
%
%
a sufficiency condition for the asymptotic stability as
%
\begin{equation}
(\alpha_{1}+\varepsilon)>|\beta_{1} f^{\prime}_{1}(x_{\tau_{1}})|.
\label{eqn1e}
\end{equation}
Note that the evolution equation of the synchronization manifold and hence the
stability condition depends 
on the $f^{\prime}_{1}(x_{\tau})$, which in turn depends on the nonlinearity of the individual systems.
From the form of the nonlinear function $f(x)$ for the MG system, 
%
%
the stability condition becomes 
%
\begin{equation}
(\alpha_{1}+\varepsilon) >\biggl|\beta_{1}\biggl(\frac{(1+x^c_{\tau_{1}})-c x_{\tau_{1}}^c}{(1+x_{\tau_1}^c)^2}\biggr)\biggr|.
\label{eqn1g}
\end{equation}
It is not possible to find the exact value of the nonlinear function $f_1^{\prime}(x_{\tau_1})$. However, 
in accordance with the Lyapunov-Razumikin  function, a special class of the Krasoviski-Lyapunov theory,
one can find the value of $f_1^{\prime}(x_{max})$ from the maximal value of $x(t)$
and arrive at an sufficient condition for  CS (GS). 

From the hyperchaotic attractor of the Mackey-Glass system for
the above chosen parameter values one can find numerically the maximum value 
of $x_{max}\approx 1.24$. 
So the stability condition becomes 
$\varepsilon>|\beta_{1} f_1^{\prime}(x_{max~\tau_1})|-\alpha_1\approx 0.33$ (which is 
a sufficiency condition for global GS). From Fig.~\ref{mgth_2c},
the threshold value of the coupling strength to attain GS in the Mackey-Glass 
systems is $\varepsilon^{(2)}_{c}\approx 0.5~(>0.33)$
for which the stability condition is satisfied.
With a similar procedure, the stability condition for the 
PWL systems become 
$(\alpha_{2}+\varepsilon)>|\beta_{2} f_2^{\prime}(x_{\tau_{2}})|$.
%
%
Now from the form of the piecewise linear
function $f_2(x)$, we have 
%
\begin{eqnarray}
f_2^{\prime}(x_{\tau_{2}})=
\left\{
\begin{array}{cc}
             1.5,&  0.8 \leq |x| \leq \frac{4}{3} \\
             1.0,&  |x| < 0.8 \\
         \end{array} \right.
\label{eqn1h}
\end{eqnarray}
Consequently, the stability condition becomes $(\alpha_{2}+\varepsilon)>|\beta_{2}|$~\cite{fn4}, 
for the asymptotic CS state $\Delta_{2}=0$.
From Fig.~\ref{mgth_2c}, we find that CS between the piecewise linear 
systems ($x_{2}$ and $x^{\prime}_{2}$) occur for the coupling strength 
$\varepsilon^{(1)}_{c}>0.27$, which satisfies the sufficient stability 
condition for  partial GS $\varepsilon>|\beta_{2}|-\alpha_{2}=0.2~(0.27>0.2)$.

%
%
Next, we demonstrate the 
existence of GS in four (only in $N=4$ for clear visibility of figures depicting
synchronization transitions) mutually coupled distinctly nonidentical
time-delay  systems in a ring, an array, a global and a star configuration. 
In addition to the above two time-delay systems, now we  have considered the
third system as a threshold piecewise linear time-delay (TPWL) 
system~\cite{dvs2010} with the nonlinear function 
%
\begin{equation}
f_{3}(x) = AF^{*}-Bx.
\label{eqn5}
\end{equation}
Here
\begin{eqnarray}
F^{*}=
\left\{
\begin{array}{cc}
-x^{*},&  x < -x^{*}  \\
            x,&  -x^{*} \leq x \leq x^{*} \\
            x^{*},&  x > x^{*}. \\ 
         \end{array} \right.
\label{eqn5a}
\end{eqnarray}
The system parameters are $\alpha_{3}=1.0$, $\beta_{3}=1.2$, 
$\tau_{3}=7.0$, $A=5.2$, $B=3.5$
and $x^{*}=0.7$ and the system exhibits hyperchaotic attractor with four positive LEs ($D_{KY}=8.211$).
As a fourth system, we take the Ikeda time-delay system~\cite{kihd1980}
with the nonlinear function $f_{4}(x) = sin(x(t-\tau_{4}))$. This system
exhibits hyperchaotic attractor with five positive LEs ($D_{KY}=10.116$)~\cite{ml2011} for 
$\alpha_{4}=1.0$, $\beta_{4}=5.0$ and $\tau_{4}=7.0$.
The schematic diagram of four mutually coupled main and their auxiliary
systems in a ring configuration is depicted in Fig.~\ref{fig1}b.
CC, CPR and MTLE for mutually coupled ring of the above four distinctly nonidentical
systems are shown in Figs.~\ref{4cring}a and ~\ref{4cring}b.  
$\lambda^{(i)}_{MTLE}>0$ and low values of $CC_{i,i^{\prime}}$  and $CPR_{i,i^{\prime}}$
for $\varepsilon=0$ indicates the main and their auxiliary systems
evolve independently. Increasing $\varepsilon$ results in decreasing 
$\lambda^{(i)}_{MTLE}$ and increasing $CC_{i,i^{\prime}}$  and $CPR_{i,i^{\prime}}$.
The PWL systems with $3$ positive LEs become synchronized first at $\varepsilon^{(2)}_{c}=0.088$
as evidenced by $\lambda^{(2)}_{MTLE}<0$, indicating the onset partial GS,
and then the MTLE of MG with $2$ positive LEs becomes negative ($\lambda^{(1)}_{MTLE}<0$) at 
$\varepsilon^{(1)}_{c}\approx 0.092$ while the other two systems are not yet synchronized.
Further, increase in $\varepsilon$ leads to CS of Ikeda systems with $5$ positive LEs at
$\varepsilon^{(4)}_{c}\approx 0.65$ with $\lambda^{(4)}_{MTLE}<0$ and 
finally the TPWL system becomes synchronized at $\varepsilon^{(3)}_{c}\approx 0.75$, confirming the
existence of a global GS between the four mutually coupled distinctly 
nonidentical time-delay systems in a ring configuration.
In addition, all $CC_{i,i^{\prime}}$ and $CPR_{i,i^{\prime}}$ reach unity (Fig.~\ref{4cring}b)
exactly at the threshold values $\varepsilon^{(i)}_{c}$ of $\lambda^{(i)}_{MTLE}$
corroborating the simultaneous existence of GS and PS.

\begin{figure*}
\centering
\includegraphics[width=1.9\columnwidth]{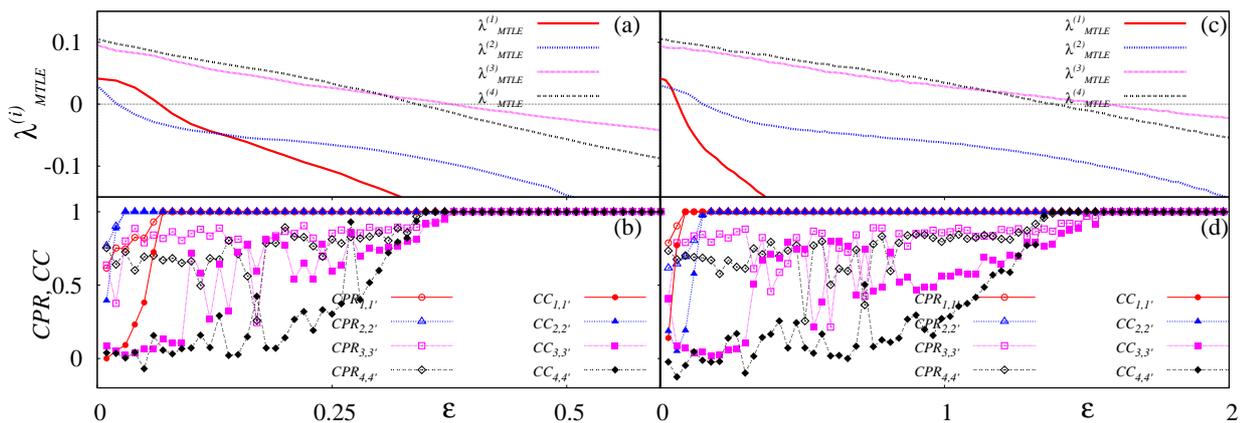}
\caption{\label{4cstar}(color on-line) (a) and (c) MTLEs and 
(b) and (d) CC, CPR of the main and auxiliary systems
of four distinctly nonidentical time-delay systems with global and star 
configurations, respectively, as a function of $\varepsilon$.}
\end{figure*}
%
%
%
Next, we illustrate the transition from partial GS to global GS in a linear array of
$N=4$ (MG, PWL, TPWL and Ikeda) systems. MTLE, CC and CPR of all the four main and their
corresponding auxiliary systems are shown in Figs.~\ref{4cring}c and ~\ref{4cring}d. 
With out coupling, all the systems evolve independently as evidenced 
by the $\lambda^{(i)}_{MTLE}>0$  and
low values of $CC_{i,i^{\prime}}$  and $CPR_{i,i^{\prime}}$. 
Again, all the four
systems reach their CS (GS) manifolds for the threshold
values $\varepsilon^{(i)}_{c}\approx 0.09, 0.04, 0.54, 1.2$, respectively,
as indicated by the transition of their $\lambda^{(i)}_{MTLE}$ below zero.
Further, GS and PS also occur together as indicated by 
$CC_{i,i^{\prime}}$ and $CPR_{i,i^{\prime}}$ 
(Fig.~\ref{4cring}d) of the systems in the array.

Maximal TLE's and CC and CPR of all the four systems with global and star
configurations are depicted in Figs.~\ref{4cstar}a, ~\ref{4cstar}b and ~\ref{4cstar}c, ~\ref{4cstar}d, respectively.
As in the case of ring configurations, $\lambda^{(2)}_{MTLE}$ transits first from
positive to negative values elucidating  CS between PWL and its auxiliary system
thereby indicating the onset of partial GS. This is followed by the MG systems, then by
Ikeda systems and finally the TPWL systems for appropriate $\varepsilon^{(i)}_{c}$ confirming the
existence of GS in four distinctly nonidentical time-delay systems with different
fractal dimensions in both global and star configurations. 

Further, we have also confirmed that there exists transition from a partial to
global GS in other permutations on the order of systems
between MG, PWL, TPWL, Ikeda systems in array and star configurations. Following
a similar stability analysis as above for the four coupled time-delay systems
with their auxiliary systems for CS,
one can also arrive at a sufficiency stability conditions for GS in all the above
four configurations. All the above results have also been confirmed in a
larger ensemble $(N=7,10)$ of time-delay systems with distinct fractal dimensions
in all the four configurations.

Finally, we have also confirmed the existence of the above transitions 
in time-delay systems with different orders, namely  
(i) in a system consisting of a mutually coupled Ikeda time-delay
system (which is a scalar first order time-delay system) and 
a Hopfield neural network~\cite{jjh1982,ml2011}
(which is a second order time-delay system), and 
(ii) in a system of mutually coupled Mackey-Glass time-delay
system (which is a scalar first order time-delay system) 
and a plankton model~\cite{era1998} (which corresponds to 
a third order system with multiple delays). Complete details
on these results will be presented in a forthcoming paper.

In summary, we have found that there exist smooth (as evidenced by
the smooth convergence of MTLE) transformation functions even for distinctly
nonidentical time-delay systems with different fractal dimensions and different orders in 
symmetrically coupled network (ring, array, all-to-all
and star), which map each of them to a common synchronization (GS) manifold.
We have also shown that the asymptotic stability of each of the system 
in the network guarantees the existence of GS as confirmed by
their MTLEs and analytical stability conditions. 
Further, we have confirmed the existence
of GS using the mutual false nearest neighbour method in all the cases (the details will be published separately).
In addition, we find
that GS always coexists with PS or vice versa in these systems using CC and CPR
and hence our understanding on the evolutionary mechanism including their dynamical
and functional behavior of systems with different fractal dimensions~\cite{csmgr1998,sff1998,pvldg2009,bbah1999,reagr2006,btgonb2000,ksat2011}
should be revisited and improved. For instance, GS leading to pathological
disorders  such as epilepsy, Parkinson's disease, paralysis, etc., may be effectively controlled with extremely small external perturbations
to appropriate variables (to electrodes) without harming the subject.
In this connection, investigations on desynchronization mechanisms and conditions
in ensembles of such systems will also be crucial.
In a recent work, GS among remote (identical) systems in simple network motifs
with a chain of delay-coupled relay elements mediating them to maintaining total
propagation delay time is 
demonstrated  using the auxiliary system approach~\cite{mcsgvs2012}. Here, the issue
of indirect connections and synchronization among remote elements in networks is 
addressed.  But we have considered different time-delay systems with different
number of positive Lyapunov exponents in regular networks and explore local
and global GS using the auxiliary system approach.
The concept of {\it equivalence} and {\it predictability}, which is well understood 
in unidirectionally coupled nonidentical systems~\cite{lkup1996},
remains an open problem for bidirectionally coupled systems. Further,
one can extend the present work to complex networks and also include heterogenity
by weighted couplings, etc., to generalize much more.

The work of R.S. and M.L. has been supported by the Department of Science 
and Technology (DST), Government of India sponsored IRHPA research project.
M.L. has also been supported by a DST Ramanna project and a DAE Raja Ramanna Fellowship.
M.L. also acknowledges the support by the Alexander von Humboldt Foundation to visit PIK,
where the work was completed.
D.V.S. and J.K. acknowledge the support from EU under project No. 240763 PHOCUS(FP7-ICT-2009-C).

\end{document}